\documentstyle[10pt,aaspp4]{article}
\def\HII{H\,{\sc ii}\ }
\slugcomment{Accepted for publication in the Astrophysical Journal}

\lefthead{Roberts, Romani, Johnston, and Green}
\righthead{The `Rabbit'}

\begin{document}

\title{The `Rabbit': A Potential Radio Counterpart of GeV J1417-6100}

\author{Mallory S.E. Roberts, Roger W. Romani}
\affil{Department of Physics, Stanford University,
    Stanford, CA 94305}

\author{Simon Johnston}
\affil{Research Centre for Theoretical Astrophysics, University of 
Sydney, NSW 2006, Australia}

\and
\author{Anne J. Green}
\affil{University of Sydney, NSW 2006, Australia}

\begin{abstract}
We have mapped the radio emission in the error ellipse of GeV J1417-6100
(2EGS J1418-6049) at 13cm and 20cm using the Australia Telescope
Compact Array. We find a large shell with extended wings , at the
edge of which is a non-thermal, polarized structure with a center
filled morphology (the `Rabbit'), coincident with an extended, hard X-ray
source. We discuss the various sources 
seen within the ellipse as potential counterparts of
the $\gamma-$ray source. We conclude that the most likely scenario
is that the Rabbit is a wind nebula surrounding a radio-quiet $\gamma-$ray
pulsar.  
\end{abstract}


\section{Introduction}

The nature of the unidentified galactic plane $\gamma-$ray sources
is one of the outstanding mysteries of high energy astrophysics.
There are $\sim$40 sources that have been detected by EGRET at
$b < 10^\circ$, many with significant flux above 1 GeV.
The only known persistent, discrete sources of GeV emission which have
firm identifications are blazars
and young pulsars. In addition, several EGRET sources are coincident with
young supernova remnants. These source types are moderately bright
sources of hard X-rays, and in most cases have associated radio emission as
well. We argue that any source drawn from the identified classes can be detected
with X-ray/radio observations.
Any error box which does not contain a source consistent
with the known GeV source types would then be a strong candidate for a new 
source class. 
Finding low energy counterparts has
been complicated by the complex $\gamma-$ray background and the high absorption
in the regions of star formation where these objects are found.
To minimize the effects of the gas and dust in these regions, it is best to
focus on energies above 1 GeV in $\gamma-$ rays, above 2 keV in X-rays,
and in the radio regime. We have begun a campaign to characterize these
sources by taking the best positions determined from the $> 1$ GeV
EGRET photons (eg. \cite{lm97}), imaging them in hard X-rays with ASCA, 
and following up with radio observations. 

The only GeV survey objects in the galactic plane with firm
identifications are 5 young pulsars, one of which, Geminga, is radio quiet.
Models of the beaming geometries required to reproduce the observed multi-
wavelength pulse profiles suggest that most of the GeV sources in the plane
are radio-quiet young pulsars as well (\cite{yr97}).  
A firm identification 
as a pulsar requires the detection of a pulse, which is not possible without
{\it a-priori} knowledge of the pulse period 
(and even then is difficult), given
the limited $\gamma-$ray data available. However, the known pulsars have 
been well studied at many wavelengths, showing properties which allow the
identification of promising candidates. 
In particular, X-ray images of young pulsars show regions of extended hard
(power law photon indices $\Gamma \sim 1-2$) emission that may
be synchrotron nebulae excited by the pulsar wind (\cite{k97}). 
At radio wavelengths, 3 of the 5 GeV pulsars (Crab, Vela, and PSR B1706-44),
have associated Pulsar Wind Nebulae (PWN), which are characterized
by radio spectral indices $ -0.3\la \alpha_R \la -0.1$ 
($F_\nu\propto \nu^{\alpha_R}$), 
amorphous morphologies, and
high fractional polarizations ($\sim 10-30 \%$) (\cite{fs97}).   
Sources with these X-ray and radio properties within GeV error ellipses
are prime candidates to be $\gamma-$ray pulsars. 
The combination of X-ray and radio images of GeV error ellipses can 
also be used to screen out the only other confirmed class of persistent
GeV point source emission, blazars. Since these active galactic nuclei
both emit hard x-rays 
($\Gamma \sim 1.7$, $F_x \ga 10^{-13}{\rm ergs\ s^{-1} cm^{-2}}  $; \cite{smu96})
and are radio-loud ($F_r \ga 0.5$ Jy; \cite{m97}),
it is relatively simple to check the possibility of a particular
low latitude source being a background blazar.

GeV J1417-6100 (\cite{lm97}, also 2EGS J1418-6049) was first recognized 
as a separate source in the supplement to the second EGRET catalog 
(\cite{t95}), 
having previously been confused with the nearby, softer source 3EG 
J1410-6147 (\cite{h98}).  X-ray observations of this region (Roberts and Romani 1998, 
hereafter RR98)
with the ASCA telescope revealed 
three hard sources X1, X2, and X3 (ps1, ps2, and ps3 in RR98) 
as well as diffuse emission
above 2 keV (Figure 1).  One of these sources (X1) is extended 
and well-centered within the GeV error ellipse; RR98 proposed it 
to be a plerion candidate, and the likely $\gamma-$ray counterpart. 
The other two appear point-like, with one (X2) possibly having 
short-term variability.  

In the radio, the low resolution 5GHz Parkes survey image shows 
a ring structure 
with a bright source (G313.5+0.2) at the northern end which has been identified
as an \HII region at a probable distance of 13.4 kpc (\cite{ch87}).   
The IRAS $60 \mu$m image shows both the ring and the \HII region.
A higher resolution image at 843 MHz, made with the Molonglo 
Observatory Synthesis Telescope (MOST) during a survey of the 
galactic plane (\cite{g94}) resolves the structure of the ring and also
shows 
wings that are coincident with the diffuse X-ray emission. There is a 
bright radio enhancement at the location of the plerion candidate, X1. 
There is no source at the position of the X-ray source X2, while
there is a faint ($\sim 7$ mJy) point source consistent with the 
position of X3, leading RR98 to conclude that they both have properties
consistent with background Seyfert galaxies.
In this paper we report on synthesis images at 20 and 13 cm
of the ring and plerion candidate made with the Australia Telescope 
Compact Array (ATCA). 
 
\section{Observations}

The ATCA consists of 6 22m dishes along an East-West baseline;
5 of the dishes are along a 3 km track, with the 6th antenna fixed a further
3km west of the track, allowing a maximum baseline of 6km.
We observed with the 1.5A (1/11/98), 0.75A (4/22/98), and 0.375 (3/28/98)
configurations.
The ATCA can observe 2 frequency bands simultaneously. For the 1.5A array, 
the central frequencies were 1344 MHz and 2496 MHz, while for the other 
2 configurations they were 1384 MHz and 2496 MHz. The bandwidth of each
was 128 MHz.
Full Stokes parameters are recorded at each wavelength.
In order to image the entire complex and to ensure that the strong \HII 
region at the edge of the ring was well within the main beam, a
mosaic with 2 pointings was done, with pointing centers near the plerion
candidate (14h 18m 40s, $-61^{\circ} 01^{\prime} 00^{\prime \prime}$[J2000]) 
and the \HII region 
(14h 19m 10s, $-61^{\circ} 51^{\prime} 00^{\prime \prime}$[J2000]).
PKS B1934-638 was used for absolute flux calibration, while PKS B1329-665 was 
used as a phase calibrator and to calibrate the antenna gains as a function
of time. Data reduction and analysis of the
data were done using the MIRIAD package (\cite{sk98}). 

The field has also been observed a second time by MOST, at 843 MHz with a 
3 kHz bandwidth, 
as part of a new galactic plane survey using a wider field of view 
($2.7^{\circ}$, \cite{g97}). This has resulted in an improved image 
without the artifacts of the original survey image caused by proximity
to the field edge (figure 1). 

\vspace*{14.3cm}
\includegraphics{fig1.ps}

\figcaption[fig1.PS]{MOST 843 MHz image of region from the new survey
(\cite{g97}) overlain with ASCA GIS 2-10 keV contours, and the EGRET $> 1$GeV 
error ellipse (RR98). The radio data show
a complex region with many unassociated sources superimposed
in the $\sim 15$kpc path through the plane. The SNR G312.4-0.4 is nearby at
a $\Sigma$-D distance of 1.9kpc; several young stellar associates are
also at similar distances (\cite{yr97}).}

\section{Results}

\subsection{Continuum Images}

Total intensity images at 20cm (figure 2) and 13cm were made without using the 
6th antenna's baselines, for a longest baseline of 1.5 km, in order to 
have reasonably complete UV coverage.
The inversion and cleaning of the data was done using the mosaicing
routines in MIRIAD, which use a maximum entropy method to remove artifacts
(\cite{s96}).
The baseline weightings were chosen to minimize sidelobes while maintaining a 
high dynamic range.
In addition, high resolution maps (figure 3) were generated by including the 6th
antenna for a maximum baseline of 6km, with uniform weighting chosen to 
maximize point source sensitivity.
All the images show a large (peak contour $\sim 12^{\prime}$ in diameter), 
circular
shell with a broad wing (K2) extending to the northeast and a narrower
wing extending to the southwest (K4). An additional spur of emission
stretches to the northwest, connecting to a bright, unresolved
source (G313.2+0.3). The entire complex reminds us of an overweight bird; since
it was observed in Australia, we refer to it as the Kookaburra. 

\vspace*{10.2cm}
\includegraphics{fig2.ps}

\figcaption[fig2.PS]{20cm ATCA image showing the `Kookaburra',
the `Rabbit' and a number of apparent background
sources. Only baselines $\le 1.5$km have been included.
Restored beam $49.1^{\prime \prime} \times 39.8^{\prime \prime}$, rms noise
level of 0.5 mJy.
}

At the north end of the shell is the
\HII region (G313.5+0.2), which is seen to be compact 
($\sim 25^{\prime \prime}$).
The high resolution images resolve this source for the first
time.  Near the center of the shell is another
source (G313.4+0.1) which the high resolution images show to be a 
typical extra-galactic core-jet. The shell itself is 
brightest in the northern half, except for a long, $\sim 8^{\prime}$, 
bright arc (K1) near the southern edge.  Our plerion 
candidate is clearly seen as a diffuse, rabbit-shaped
structure at the west edge (figure 2). 
In the MOST image and the 20cm map there can be seen an additional 
diffuse shell to the southwest of the Kookaburra (G313.0-0.1).
Due to the smaller
primary beam, this is out of the field of view of the 13cm images. 
The MOST image also includes the nearby supernova remnant G312.4-0.4 (figure 1).

\vspace*{12.0cm}
\includegraphics{fig3.ps}

\figcaption[fig3.PS]{High resolution 13cm image of part of the shell, with
$2.9^{\prime \prime}\times 2.7^{\prime \prime}$ beam, rms noise = 0.2 mJy/beam 
made with uniform weighting 
of the longest baselines. Contour levels of 8, 12, and 20 mJy/Beam are from 
the 13cm, baseline $\le$1.5 km, $31.8^{\prime \prime}\times 25.9^{\prime \prime}$ beam 
maps which better preserve flux. Positions of the 3 X-ray 
sources are indicated. The \HII region and central source are
clearly resolved.  No unresolved
point sources are seen in the Rabbit.} 

\subsection{Spectral Index Studies}

With a relatively sparse array like the ATCA, differing UV coverage
at different frequencies can lead to incorrect determination of spectral
indices if only the total fluxes of extended regions are used.
Gaensler et al. (1998a) discuss this problem and have 
developed a method to obtain fairly reliable spectral 
measurements with ATCA and MOST observations. This method consists
of spatially filtering observations at different frequencies so that
their effective UV coverage is identical.
Images are then produced with identical
beam sizes and resolutions of $\sim 1$ pixel/beam. 
Several adjacent pixels are used for flux comparison. If the region
has a constant spectral index, pixel values from images of different
wavebands plotted on a temperature-temperature
(T-T) plot (\cite{t62}) should well fit a line $F_1 = aF_2 +b$. The  
slope of this line is used to calculate the spectral index ($\alpha_R = 
{\rm log} a /{\rm log} [\nu_1 / \nu_2]$) in order to account for any
systematic offset in the flux levels.  Such a plot for the
Rabbit is shown in figure 4. Using this method, 
we have measured spectral indices between the 843 MHz, 1384 MHz,
and 2496MHz spectral bands for various parts of the field.  
In order to better constrain the fits, only pixels with 20cm 
flux densities above 20mJy/beam for region K2 and 12 mJy/beam for K4 were
included. Spectral indices
are summarized in table 1. Due to the smaller primary beam size, the 13cm
maps had problems with artifacts at the edges of regions K2 and K4, so these 
areas were only fit between the 20cm and 36cm bands. 

\begin{deluxetable}{lccccc}
%\footnotesize
\tablecaption{Spectral Index Measurements \label{tbl-1}}
\tablewidth{0pt}
\tablehead{
\colhead{Region} & ID & $F_{20cm}$ (mJy)& \colhead{$\alpha_{20/36}$} 
& \colhead{$\alpha_{13/36}$} & \colhead{$\alpha_{13/20}$} 
} 
\startdata
Rabbit& PWN & 400 & $-0.25\pm 0.02$&$-0.25\pm0.06$&$-0.33\pm 0.08$\nl
G313.5+0.2& \HII & 650 & $0.7\pm 0.3$&$0.5\pm0.1$&$0.3\pm0.1$\nl
G313.4+0.1& Ex. Gal. & 50 &$-1.2\pm 0.3$&$-1.2\pm0.1$&$-1.0\pm0.3$\nl
G313.2+0.3& Ex. Gal. & 350 & $-1.2\pm 0.4$&$-1.0\pm0.2$&$-0.8\pm0.2$\nl
K1& (arc) & 800 &$0.2\pm0.1$&$0.1\pm0.1$&$-0.1\pm0.1$\nl
K2&&1000&$-0.2\pm0.2$&---&---\nl
K3&&20&$-0.4\pm0.5$&$-0.4\pm0.3$&$-0.5\pm0.3$\nl
K4&&400&$-1.2\pm0.5$&---&---\nl
\enddata
\end{deluxetable}

The bright point source G313.2+0.3, the double source
near the center of the shell, and the point source associated with
the X-ray source X3 from RR98, all are too small to have a 
significant number of pixels to fit in the above manner. We therefore
measured their fluxes directly from the high resolution maps. Since these 
sources are mostly unresolved, their flux levels should not be affected 
by differing UV coverage.

\vspace*{7.8cm}
\includegraphics{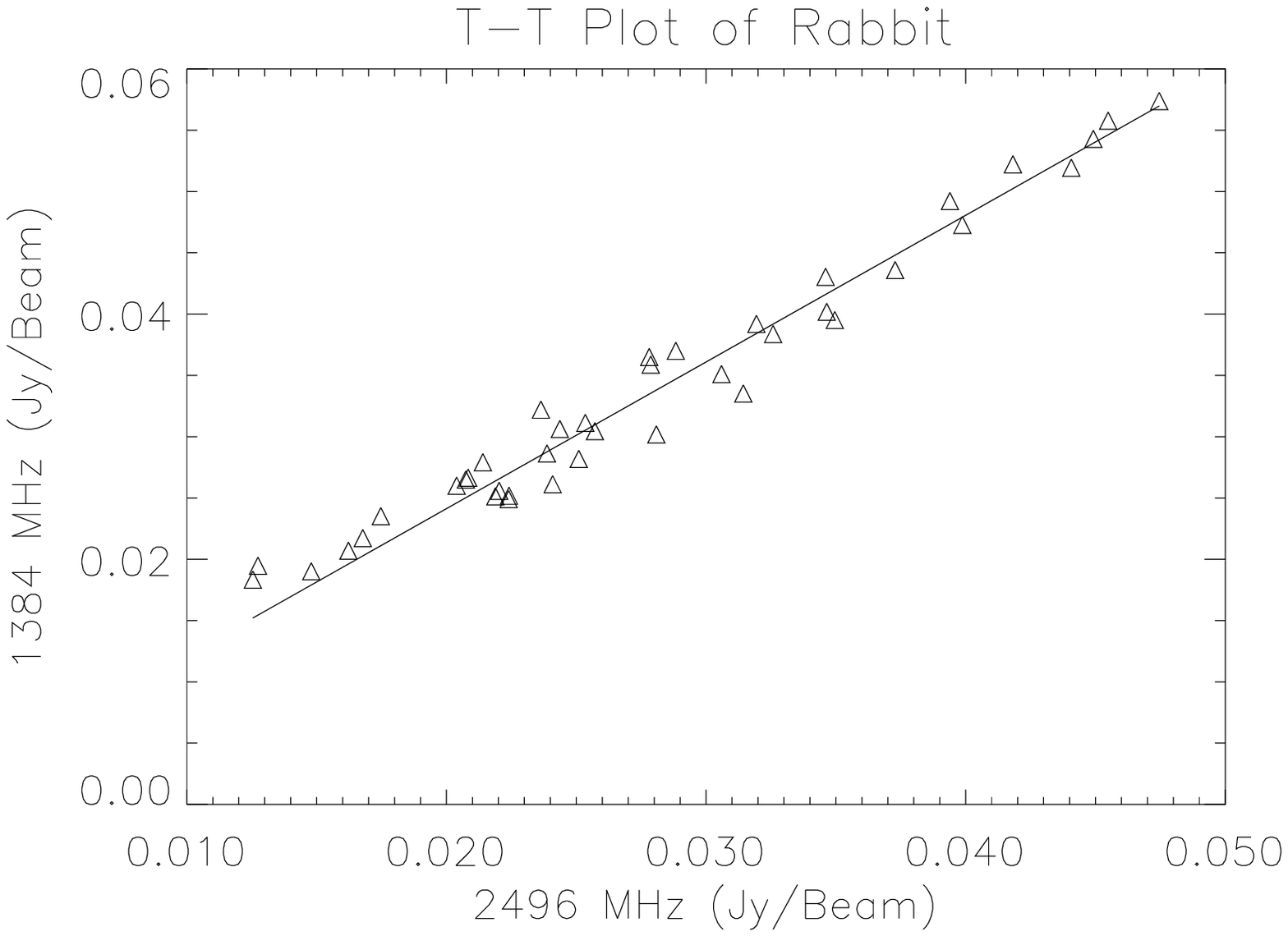}

\figcaption[fig4.PS]{Sample T-T plot, showing the spectral fit of the
Rabbit to the 20cm and 13cm maps.}

\subsection{Polarization Studies}

Since the ATCA simultaneously observes both linear polarizations,
full Stokes parameter maps were obtained for all frequencies and
configurations. 
Each 128 MHz observing band consists of 32 4MHz
channels, which are combined in the processing into 16 8MHz bands,
the central 13 of which (except where interference
prevents this) may be used to create linear polarization maps.
Since bandwidth depolarization across 128 MHz can be a serious problem, 
especially at 1400 MHz, the following procedure was done.
For each channel, Q and U maps were generated and minimally CLEANed for
each of the two pointings, using all 3 configurations but excluding 
antenna 6. The two pointings were linearly mosaiced together for the full
Q and U maps. These were then combined to make total polarization maps and
polarization angle maps of each 8 MHz channel. All 13 of the 13cm channels
were averaged together to make the total linear polarization maps. Due to
the different central frequency of the 1.5 km configuration, there were
only 8 overlapping channels at 20cm to average.
The polarization angle maps were fit pixel by pixel in
order to generate rotation measure maps. Pixels that could not be well
fit were blanked.

In order to determine fractional polarization, total intensity maps were 
generated using the same inversion, CLEANing, and linear mosaicing parameters
as the polarization maps, but using the full 13 channel bandwidth. 
The final polarization maps were then divided by these total intensity
maps. 
The noise level was relatively high, so pixels which were not 
$\sim 3 \sigma$ above background in both polarized intensity and total
intensity were blanked to avoid spurious measurements of high polarization.
This means there may be regions of high fractional
polarization but low total intensity that were not included in the
maps. It should be noted that, since the ATCA simultaneously observes at
2 different frequencies using a combination feed, only one of the
frequencies is at the optimum focus, in this case the 20cm band. This
means that the 13cm polarization maps are more susceptible to instrumental
polarization leakages for bright sources.
This is evidenced by significant polarized flux coming from the \HII region
in the 13cm maps.
Therefore, we use the 20cm maps for quantitative analysis, although we note
that morphological features seen in the 20cm total polarization map 
are reproduced in the 13cm map, and that the apparent fractional polarization
is higher at 13cm. 

\vspace*{11.3cm}
\includegraphics{fig5.ps}

\figcaption[fig5.PS]{20cm fractional linear polarization (greyscale) with
a polarized flux cut of 0.8mJy/beam and total flux cut of 3.5mJy/beam ($28.2^{
\prime \prime}\times 25.5^{\prime \prime}$ beam), with 20cm total 
flux contours of the Rabbit at 3.5, 7, 10.5, and 14 mJy/beam superimposed. Fractional polarization reaches 
$\sim 15$\% in the
`paw' regions P1 and P2. A spur stretching to the NW also has high
polarization; residual polarization in the rest of the image is quite
low. The circle marks the core
position of the {\it ASCA} x-ray emission.}

\section{Discussion}

\subsection{Multi-Wavelength Comparisons}

We now consider the morphological, spectral, and polarization properties
of the individual sources in the field. 
	
The \HII region (G313.5+0.2) is seen to be small, with a linear size of $\sim 1.5$ pc, and
a total 20cm flux of $\sim 650$ mJy. 
It has two central peaks, separated by $\sim 0.4$pc, and no evidence of a 
shell component (figure 3). 
The spectrum does not seem to have turned over to the 
optically thin regime at our observed frequencies.
This source is quite bright in the IRAS image, with 
the ratio of 60 $\mu$m to 843 MHz flux density $R_{IR/R}\sim 2000$. 
The small size, high turnover frequency, high $R_{IR/R}$, plus 
the observation of associated ${\rm H_2O}$ and OH masers 
(\cite{c89}), suggest
that this is a
$compact$ \HII regions (\cite{g88}, and references therein); such sources may be involved with the 
early stages of star formation.  

The bright point source G313.2+0.3 is unresolved, with a 20cm flux of
$\sim 350$ mJy, and has a steep spectrum
($\alpha \sim -1$). There is no evidence of associated X-rays in the ASCA
image, and no infrared excess in the $60 \mu$m IRAS image.  
We therefore identify it with the class of extra-galactic compact steep spectrum
radio sources which make up a large fraction of radio point
sources (\cite{o98}). The double source near the center of the Kookaburra (G313.4+0.1)
has the typical morphology of a radio galaxy,
a steep spectrum, 20cm fluxes of $\sim 25$ mJy and $\sim 29$ mJy 
for the two components, and no evidence of X-ray or 60 $\mu$m emission.

The radio source coincident with the hard X-ray source X3 is unresolved 
(figure 3). We measure an improved (J2000) position of (14h 18m 37.7s,
$-60^{\circ} 45^{\prime} 00^{\prime \prime}$).
Its spectrum appears flat between 20cm and 13cm, with a flux of $\sim 1.8$ mJy. 
The MOST 843 MHz flux varied between the first and second survey
images, changing from $\sim 7$ mJy to $\sim 0.3$ mJy. 
The variability and overall flux are consistent with its identification
as a radio-quiet AGN. Upper limits of $\sim 0.5$ mJy at 20cm and
$\sim 0.2$ mJy at 13cm can be
placed on any point source emission associated with X2 of RR98.

The Kookaburra as a whole does not have uniform properties.
The diffuse X-rays seem to stretch from just SW of the Rabbit up into 
the NE wing (region K2). Unfortunately, the southern arc (region K1) and
much of region K4 are either outside or right at the edge of the
ASCA GIS field of view. However, the observed X-rays do not appear to
be tracing out the radio shell which makes up the body of the
Kookaburra, suggesting that the wings and the shell may be
unrelated. If this is so, K1 would
be the best region to examine to determine the character
of the shell. The fit spectra are consistent with thermal emission, although
the scatter in the T-T plot may indicate a non-thermal
component to the shell. However, taking into consideration the associated 
$60 \mu$m flux seen by IRAS ($R_{IR/R}\sim 500$), it is likely that the shell's emission
is predominantly thermal in nature. Assuming the \HII region, the Rabbit, 
the central source, and the wings are not part of the shell, we estimate 
the total 20cm flux to be $\sim 6.5$ Jy.  

Region K2 has a rectangular shape with a filamentary sub-structure, 
and $\sim 1$ Jy of total 20cm flux. The base of this filament (K3) 
is somewhat brighter with $\sim 20$ mJy in excess of the surrounding
wing, and coincident with a hard, bright knot in X-rays. 
The spectrum of K2 
as a whole is not well constrained, having a large spread in its T-T
plot. The analysis is complicated as well by the smaller beam size at 13cm, 
limiting the
northern regions to comparisons only between the 20cm map and the MOST image.
Spectral measurements of K3, while having large
uncertainties, are steeper than expected from a thermal source.
The non-thermal nature of this area is supported by the hard X-ray spectrum 
and a lack of correlated $60 \mu$m emission.
Region K4 also shows filamentary structure. It has
an overall lower intensity than K2, with a total flux of
$\sim 400$ mJy, making it more
difficult to measure its spectrum. Again we find no correlation with the 
IRAS flux in this region.
In the 20cm polarization map and rotation measure map, an
excess of pixels in the wings (K2 and K4) survive our goodness of fit cuts, 
suggesting
they may be significantly polarized, although the overall low signal to
noise prevents accurate estimates of the polarized flux.

The low resolution images of the Rabbit show it to have a 
centrally peaked morphology, with the peak of the X-ray emission coincident
with the front ``paw" region P1, 
and a total 20cm radio flux of $\sim 400$ mJy. 
A direct comparison of the X-ray morphology to the radio
is difficult given the broad ASCA PSF and the
extreme distortion when it is near the edge of the ASCA GIS field of view. 
However,
it should be noted that the fit 2D Gaussian profile of the X-ray source, when
compared to that of a point source is extended in a north-south direction, as
is the Rabbit.
The high resolution radio images (figure 3)
fail to detect any unresolved sources (upper limits of $\sim 0.8$ mJy
at 20cm, $\sim 0.3$ mJy at 13cm), although there is a faint, $\sim 
6$ mJy, point source just south of the Rabbit.
The radio spectrum of the Rabbit is clearly non-thermal, 
with a spectral index of $-0.25 \pm 0.1$. 
The IRAS image shows slight evidence of increased IR emission in this
portion of the shell, with the ratio of the excess IR emission to excess radio
$R_{IR/R}\la 35$. In the 20 cm polarization map,
the Rabbit's `paws', P1 and P2, are seen to be the only sources of polarized
flux in the entire field with intensities greater than 2 mJy/beam (figure 5). 
This implies linear polarization fractions of $\sim 10\%$.
The rotation measure across the 20cm band varies with position, with
the polarization peak at P1 having RM$=-150 \pm 20$ rad 
${\rm m}^{-2}$, and the
peak P2 having RM=$-240 \pm 20$ rad ${\rm m}^{-2}$. There is
an additional spur of polarized emission extending to the northwest.
The polarization fraction at 13cm appears even higher, suggesting
significant Faraday depolarization at 20cm.

\subsection{A Radio Counterpart of GeV J1417-6100}

The plethora of radio sources within the GeV error ellipse would, at first
glance, make the identification of a low energy counterpart difficult. 
However, several of the sources (G313.4+0.1, G313.2+0.3, X3) 
are readily identified
with common extra-galactic sources which are not known to be $\gamma-$ray
emitters. This is true of the \HII region as well. Both known 
$\gamma-$ray source classes (pulsars and blazars) have hard X-ray
counterparts. This leaves the Rabbit and portions of the wings
as viable radio counterparts to the GeV source.

The nature of the body of the Kookaburra is uncertain. 
The common wisdom is that if a shell is thermal, which the spectrum of region K1 
and the infrared flux seem to suggest,  it is not a supernova remnant.
Whiteoak and Green (1996) list typical values of $R_{IR/R}$ for various objects,
and the observed value for the shell is consistent with that of a planetary 
nebulae.  The lack of a bright central
infrared or optical source argues against a massive stellar wind bubble
interpretation, although this remains a possibility.

The wings of the Kookaburra (regions K2 and K4) are at least partially 
non-thermal. While the infrared
emission in the region is substantial, its morphology does not 
correspond with the wings. 
The filamentary structure of the wings 
is not typical of a thermal shell.
One possible interpretation is that the wings are part of a large 
($\sim 2^{\circ}$) old SNR shell; in figure 1 there is faint radio emission 
continuing in an arc to the west.  The presence of X-rays and 
non-thermal radio suggest that region K2 has been re-energized
with non-thermal electrons. Region K3,
having both excess radio and hard X-rays, might be a site of nonthermal
$e^\pm$ acceleration ({\it eg.} a pulsar), but with the current
data the evidence for this is far from compelling.

The nature of the emission from the Rabbit is certainly non-thermal; 
its significant polarized flux and low IR/R
emission imply a supernova remnant identification. 
SNR with $\alpha_R \ga -0.3$, polarization fractions
$\ga 10\%$, and a filled center morphology are usually classified as plerionic, or a PWN.
Comparing the polarization fraction of P1 and P2 and the total spectral index 
with 
PWN around known $\gamma-$ray pulsars (\cite{fs97}), we see that the Rabbit 
is within the observed range of values. Although there
are no direct measurements of the distance, if we assume that it is 
related to the young objects in the region (Clust 3, Cl Lynga 2, or SNR 
G312.4-0.4, \cite{yr97} and references therein) we
infer a distance of $\la 2$kpc. 
For $\alpha_R = -0.25$, $F_{20cm} = 400$ mJy, and a distance of 2 kpc, 
the luminosity of the Rabbit integrated from $10^7 - 10^{11}$ Hz 
is $\sim 10^{31.9}$ ergs/s, which is near the lower end of the range of
other known PWN (\cite{fs97}). However, the wind nebula associated with 
PSR B0906-49 has an $L_r\sim 10^{30}$ ergs/s (\cite{g98b}), 
and the $\gamma$-ray pulsar PSR B1055-52 shows no evidence of extended
radio emission (Gaensler, private communication), 
indicating PWN luminosities can be even lower than the Rabbit's. 

\vspace*{9.5cm}
\includegraphics{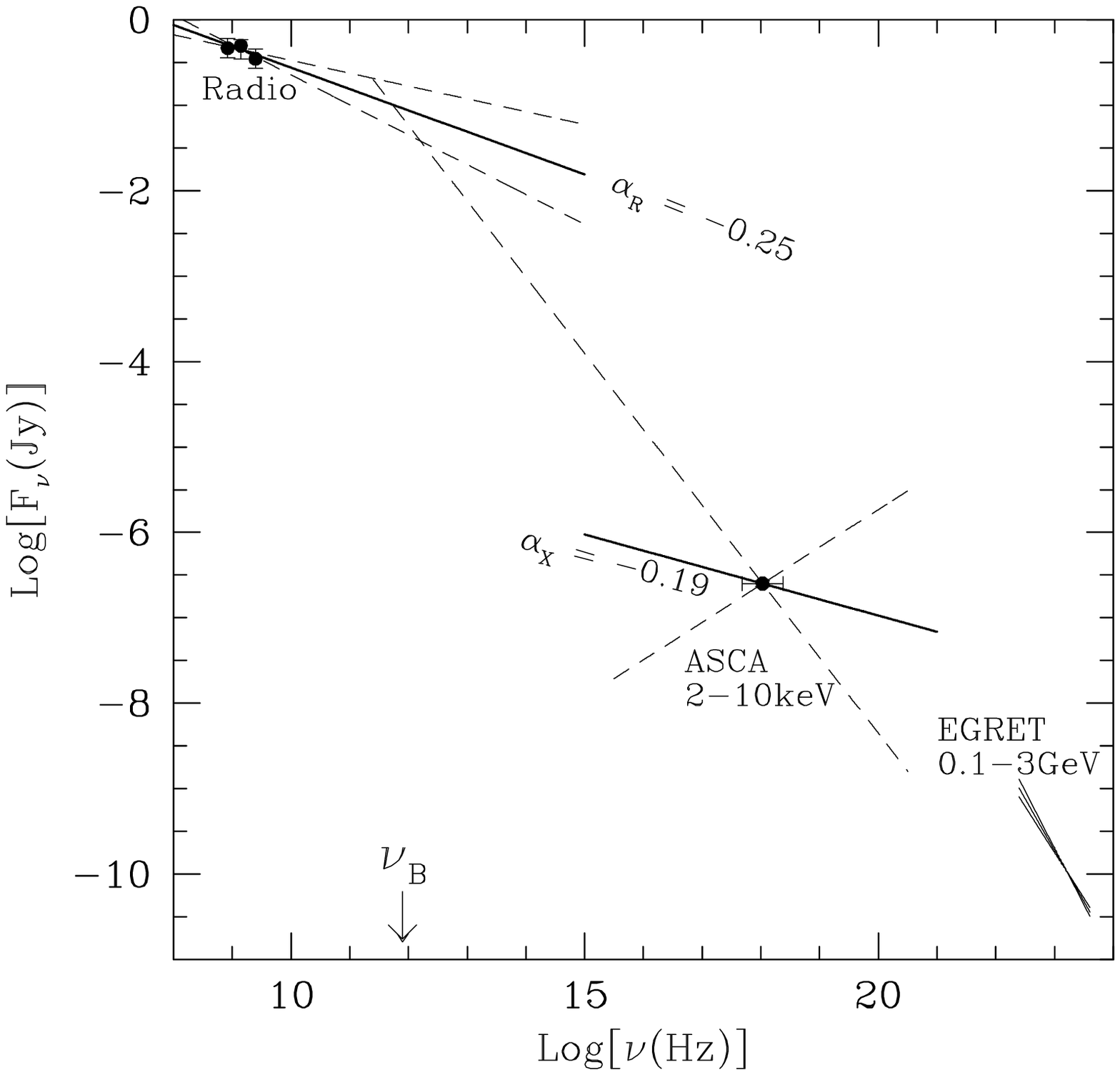}

\figcaption[fig6.PS]{Broad-band spectrum of the Rabbit. A break
$\nu_B \la 0.3$THz is required between the radio and X-ray regimes,
presumably associated with electron cooling. The {\it ASCA} determination of
$\alpha_X$ is at present quite poor. Dashed lines show bounds of the
multi-parameter 90\% errors (not one-parameter as stated in RR98), with
the long dashed line extended to indicate the upper limit to the spectral break.
The well-constrained index and flux of the GeV emission 
clearly requires an extra spectral component, plausibly emission from a young pulsar.}

One challenge in interpreting our observations rests in the 
unknown distance of several portions of the Kookaburra. The bright point source
G313.2+0.3 is of course likely in the background. If the
Rabbit is as distant as the \HII region at 13.4kpc, it's radio luminosity
would be comparable to the brightest PWN. On the other hand, if the
GeV source is associated with the Rabbit it would be some 20 times more
luminous than the Crab -- we therefore feel that the ~2kpc distance is
more plausible and the \HII region is also an unassociated background
source. The non-thermal wings and Rabbit might well be at the same distance,
but it is unclear if these are associated with the shell. HI measurements
of the shell and Rabbit would be challenging, but very illuminating in
sorting out this region's associations.

The broad band spectrum of the Rabbit is shown in figure 6. 
A spectral break or cutoff is required somewhere between the radio and 
X-ray bands. If we assume the observed radio and X-ray fluxes are 
synchrotron emission coming from a single population 
of electrons $N(\gamma ) \sim \gamma^p$, 
we expect the electron power law to steepen from
$p \approx -1.5$ to $p \approx -2.5$ above a break energy corresponding
to the synchrotron lifetime. The implied $\alpha_X \approx -0.75$ is consistent
with the present large error bars; the resulting break frequency is
$\nu_B \la 3 \times 10^{11}$Hz.
If this is the case, the {\it EGRET} photons are clearly an additional 
component, lying well
above the extrapolated radio to X-ray spectrum with a steeper spectral
index. If the X-rays are dominated by a different
electron population which is also the source of the $\gamma-$ray emission,
a spectral break is required between the X-rays and $\gamma-$rays. This 
would require the cooling break in the radio to be lower than implied
above (of course if the break
is intrinsic to the $e^\pm$ population it is at present poorly constrained).

	At present, the radio/X-ray data suggest a plerion, but do not prove
a pulsar origin of the GeV emission. However, if we adopt
the pulsar plus plerion interpretation, we can use the various spectral
components to constrain the putative pulsar's properties and to check
the global consistency of the model. For example, our estimated $\nu_B$
implies $B \sim 370 \tau_4^{-2/3}\mu$G for a nebula age of $10^4\tau_4$y.
If we assume the post-shock nebular field is contributed by a pulsar
of surface dipole $10^{12}B_{12}$G and age $10^4\tau_4$y with a toroidal
field out to a wind shock at angle $\theta_s=\theta\prime$ arcmin, 
then we expect
$B_s \sim 3B_\ast r_\ast^3/(r_{LC}^2r_s) 
\sim 230(B_{12}\tau_4d_{\rm kpc}\theta^\prime)^{-1} \mu$G with 
$r_s = 0.3 d_{\rm kpc}\theta^\prime$pc. 
Then, as RR98 noted, if we adopt the empirical PWN $L_X({\dot E_{SD}})$ law
of Kawai, et al. 1998, we infer a spindown luminosity of 
${\dot E}_{SD} \approx 3 \times 10^{35} d_{\rm kpc}^{1.6}$ erg/s which translates to the
constraint $B_{12} \approx 18 \tau_4^{-1}d_{\rm kpc}^{-0.8}$. The fiducial
pulsar values $B_{12} \approx 3$, $\tau_4 \approx 3$ satisfy these constraints
and also predict (\cite{r96}) a GeV pulsar luminosity comparable to the observed
flux of GeV J1417-6100. The wind shock field estimated from these parameters
agrees with the spectral estimate from $\nu_B$ above if 
$\theta_s$ is rather small, $\sim 5^{\prime\prime}$. Such a shock standoff
would be reasonable if the pulsar has a modest space velocity
$\theta_s \approx ({\dot E}_{SD}/4\pi n m_H c)^{1/2} d^{-1} v_p^{-1}
\approx 5^{\prime\prime} n^{-1/2} d_{\rm kpc}^{-0.25}v_7^{-1}$. Alternatively,
the low $\nu_B$ might arise if the postshock field amplifies beyond
equipartition.

The rotation measure gives us an independent estimate of $B_{neb}$, if
we assume a local electron density. The 
Galactic RM in this direction we estimate to be $\sim -100$ rad/${\rm m^2}$
at $\sim 2$kpc (\cite{rl94}), 
which is consistent with the values measured for the surviving
pixels in the wings. This gives an intrinsic RM of $\sim -50$ rad/${\rm m^2}$ 
for P1, and $\sim -140$ rad/${\rm m^2}$ for P2.
So since RM$=0.8 \int n_e B_{\parallel\mu}{\rm d}l_{pc}$ we infer an
average value $\langle n_e B_\parallel \rangle 
\sim 200 (\theta^\prime d_{\rm kpc})^{-1} \mu{\rm G\,cm^{-3}}$ in the 
nebula.

\subsection{Potential Pulsar Birthsites}

There are several potential birthsites of a pulsar in the region. One is 
the Rabbit itself, whose properties are consistent with a plerionic SNR. 
However, its angular size is rather small for the assumed age and distance.
Another is SNR G312.4-0.4, which is $\sim 63^{\prime}$ from the Rabbit. 
This SNR is notable for its relatively flat radio index ($\alpha_R \sim
-0.38$, \cite{wg96}), suggesting that non-thermal electrons were injected
into the remnant. The west half is also coincident with the softer $\gamma-$ray source
3EG J1410-6147 (figure 1). However, as noted in RR98, this site would require a high
transverse velocity $v_p=1.1\times10^3 (d_{\rm kpc}/2) (t_4/3)^{-1}$km/s,  
if the $\Sigma - D$ distance, which is often unreliable, of $\sim 1.9$ kpc
is correct. The small shell G313.0-0.1, which is only $\sim 25^{\prime}$ away,
is also a potential SNR. 
It is interesting to note that the morphology of the `paws', plus the position
of the X-ray source near this region (figure 5) suggests a bow shock,
with the body of the Rabbit a wake, implying an origin to the west. On the
other hand, a line drawn from the polarization peak P1 through 
P2 passes close to the centers of both G313.0-0.1 and 
SNR G312.4-0.4.  
 
Finally, the ``wings" may themselves represent a large, old shell birthsite.
With an age of $10^5$y, and distance of 1-2kpc, the implied transverse 
pulsar velocity
of $\sim 170 d_{\rm kpc}$ km/s is reasonable. Then, as in CTB80/PSR1951+32, the old
shell is being re-energized by PWN electrons (cf. \cite{sfs89}). 
High resolution X-ray imaging,
coupled with our radio maps will be very helpful in checking the scale and
orientation of the implied bow shock morphology.

\section{Conclusions}

We conclude that the Rabbit is the most likely counterpart of GeV 
J1417-6100, and its radio properties are consistent with it being
a pulsar powered plerion, with the most likely location
of the pulsar being near P1. The Kookaburra is a thermal
shell of unclear origin, with non-thermal wings that may
not be physically associated.  High resolution 
X-ray images would be useful in determining which features in the 
radio are not associated with the shell, searching for the magnetospheric flux
of any neutron star within the Rabbit and checking if the `paw' region has
a bow shock morphology.  Stronger limits on point sources within
the Rabbit are needed to rule out a pulsar visible in the radio, but
statistically we should not be surprised if this is a GeV pulsar with
a radio beam missing Earth. If true, this scenario will only receive
detailed confirmation with a pulse period detection at high energies.

\acknowledgments
We thank Bryan Gaensler for useful discussions on ATCA and MOST 
data analysis. The Australia Telescope is funded by the 
commonwealth of Australia for operation as a National Facility 
managed by the CSIRO. The MOST is owned and operated by the University of
Sydney, with support from the Australian Research Council and the
Science Foundation within the School of Physics. This work was supported 
in part by NASA 
grants NAG5-3333 and NAGW-4562 and a grant from the Cottrell Foundation.

\end{document}